\documentclass[apjl]{emulateapj}
\usepackage{comment}
\usepackage{ifthen}
\usepackage{lineno} 
\usepackage[fleqn]{amsmath}

\newcommand{\forloop}[5][1]%
{%
\setcounter{#2}{#3}%
\ifthenelse{#4}%
	{%
	#5%
	\addtocounter{#2}{#1}%
	\forloop[#1]{#2}{\value{#2}}{#4}{#5}%
	}%
	{%
	}%
}%

\newcommand{\ctbd}[1]{}

\newcommand{\lc}{light curve}
\newcommand{\lcs}{light curves}
\newcommand{\Lc}{Light curve}

\newcommand{\band}[1]{\ensuremath{#1}~band}

\newcommand{\kms}{\ensuremath{\rm km\,s^{-1}}}
\newcommand{\ms}{\ensuremath{\rm m\,s^{-1}}}

\newcommand{\gcmc}{\ensuremath{\rm g\,cm^{-3}}}
\newcommand{\ergscmsq}{\ensuremath{\rm erg\,s^{-1}\,cm^{-2}}}

\newcommand{\vsini}{\ensuremath{v \sin{i}}}
\newcommand{\feh}{\ensuremath{\rm [Fe/H]}}
\newcommand{\meh}{\ensuremath{\rm [m/H]}}

\newcommand{\rsun}{\ensuremath{R_\sun}}
\newcommand{\msun}{\ensuremath{M_\sun}}
\newcommand{\lsun}{\ensuremath{L_\sun}}

\newcommand{\rstar}{\ensuremath{R_\star}}
\newcommand{\mstar}{\ensuremath{M_\star}}
\newcommand{\lstar}{\ensuremath{L_\star}}

\newcommand{\teffstar}{\ensuremath{T_{\rm eff\star}}}
\newcommand{\rhostar}{\ensuremath{\rho_\star}}
\newcommand{\loggstar}{\ensuremath{\log{g_{\star}}}}

\newcommand{\rearth}{\ensuremath{R_\earth}}
\newcommand{\mearth}{\ensuremath{M_\earth}}

\newcommand{\rpl}{\ensuremath{R_{p}}}
\newcommand{\mpl}{\ensuremath{M_{p}}}

\newcommand{\rhopl}{\ensuremath{\rho_{p}}}

\newcommand{\arstar}{\ensuremath{a/\rstar}}
\newcommand{\zrstar}{\ensuremath{\zeta/\rstar}}

\newcommand{\rjup}{\ensuremath{R_{\rm J}}}
\newcommand{\mjup}{\ensuremath{M_{\rm J}}}

\newcommand{\refsecl}[1]{\mbox{Section \ref{sec:#1}}}

\newcommand{\reftabl}[1]{Table~\ref{tab:#1}}

\newcommand{\flwof}{\mbox{FLWO 1.2\,m}}

\newcommand{\hatcurhtr}{HTR220-003}                                    
\newcommand{\hatcurCCra}{\ensuremath{06^{\mathrm h}43^{\mathrm m}23.52{\mathrm s}}}                                  
\newcommand{\hatcurCCdec}{\ensuremath{+27{\arcdeg}15{\arcmin}08.2{\arcsec}}}                                 
\newcommand{\hatcurCCtwomass}{2MASS~06432353+2715082}                  
\newcommand{\hatcurCCgsc}{GSC~1901-00976}                              
\newcommand{\hatcurCCtassmv}{\ensuremath{10.908\pm0.036}}              
\newcommand{\hatcurCCtassmvshort}{\ensuremath{10.9}}                   
\newcommand{\hatcurCCtassmB}{\ensuremath{11.287\pm0.040}}              
\newcommand{\hatcurCCtassmI}{\ensuremath{9.919\pm0.093}}               
\newcommand{\hatcurCCtassmg}{\ensuremath{11.040\pm0.030}}              
\newcommand{\hatcurCCtassmr}{\ensuremath{10.762\pm0.080}}              
\newcommand{\hatcurCCtassmi}{\ensuremath{10.709\pm0.060}}              
\newcommand{\hatcurCCtwomassJmag}{\ensuremath{10.068\pm0.021}}         
\newcommand{\hatcurCCtwomassHmag}{\ensuremath{9.881\pm0.017}}          
\newcommand{\hatcurCCtwomassKmag}{\ensuremath{9.830\pm0.016}}          
\newcommand{\hatcurLCrprstar}{\ensuremath{0.10540\pm0.00086}}          
\newcommand{\hatcurLCbsq}{\ensuremath{0.7613_{-0.0104}^{+0.0077}}}     
\newcommand{\hatcurLCimp}{\ensuremath{0.8725_{-0.0060}^{+0.0044}}}     
\newcommand{\hatcurLCzeta}{\ensuremath{29.37\pm0.21}}                  
\newcommand{\hatcurLCdur}{\ensuremath{0.09463\pm0.00093}}              
\newcommand{\hatcurLCingdur}{\ensuremath{0.0336\pm0.0021}}             
\newcommand{\hatcurLCP}{\ensuremath{2.7908327\pm0.0000047}}            
\newcommand{\hatcurLCPshort}{\ensuremath{2.7908}}                      
\newcommand{\hatcurLCT}{\ensuremath{2456553.61645\pm0.00042}}          
\newcommand{\hatcurSMEiteff}{\ensuremath{6516\pm50}}                   
\newcommand{\hatcurSMEizfeh}{\ensuremath{-0.071\pm0.080}}              
\newcommand{\hatcurSMEizfehshort}{\ensuremath{-0.07}}                  
\newcommand{\hatcurSMEilogg}{\ensuremath{4.16\pm0.10}}                 
\newcommand{\hatcurSMEivsin}{\ensuremath{40.14\pm0.50}}                
\newcommand{\hatcurSMEivmac}{\ensuremath{0.0}}                         
\newcommand{\hatcurSMEivmic}{\ensuremath{0.0}}                         
\newcommand{\hatcurSMEiiteff}{\ensuremath{6566\pm50}}                  
\newcommand{\hatcurSMEiizfeh}{\ensuremath{-0.077\pm0.080}}             
\newcommand{\hatcurSMEiizfehshort}{\ensuremath{-0.077}}                
\newcommand{\hatcurSMEiilogg}{\ensuremath{4.235\pm0.041}}              
\newcommand{\hatcurSMEiivsin}{\ensuremath{40.06\pm0.50}}               
\newcommand{\hatcurLBii}{\ensuremath{0.1716}}                          
\newcommand{\hatcurLBiii}{\ensuremath{0.3692}}                         
\newcommand{\hatcurLBir}{\ensuremath{0.2396}}                          
\newcommand{\hatcurLBiir}{\ensuremath{0.3825}}                         
\newcommand{\hatcurLBikep}{\ensuremath{0.3239}}                        
\newcommand{\hatcurLBiikep}{\ensuremath{0.3235}}                       
\newcommand{\hatcurISOmshort}{\ensuremath{1.30}}                       
\newcommand{\hatcurISOmlong}{\ensuremath{1.296\pm0.036}}               
\newcommand{\hatcurISOrshort}{\ensuremath{1.43}}                       
\newcommand{\hatcurISOrlong}{\ensuremath{1.428\pm0.030}}               
\newcommand{\hatcurISOrholong}{\ensuremath{0.627\pm0.033}}             
\newcommand{\hatcurISOlogg}{\ensuremath{4.240\pm0.015}}                
\newcommand{\hatcurISOlum}{\ensuremath{3.39\pm0.19}}                   
\newcommand{\hatcurISOmv}{\ensuremath{3.411\pm0.067}}                  
\newcommand{\hatcurISOage}{\ensuremath{2.01\pm0.35}}                   
\newcommand{\hatcurISOMK}{\ensuremath{2.398\pm0.048}}                  
\newcommand{\hatcurRVK}{\ensuremath{262\pm30}}                         
\newcommand{\hatcurPPi}{\ensuremath{82.13\pm0.18}}                     
\newcommand{\hatcurPPlogg}{\ensuremath{3.402\pm0.055}}                 
\newcommand{\hatcurPPar}{\ensuremath{6.37\pm0.11}}                     
\newcommand{\hatcurPParel}{\ensuremath{0.04230\pm0.00039}}             
\newcommand{\hatcurPPrho}{\ensuremath{0.86\pm0.12}}                    
\newcommand{\hatcurPPmshort}{\ensuremath{2.18}}                        
\newcommand{\hatcurPPmlong}{\ensuremath{2.18\pm0.25}}                  
\newcommand{\hatcurPPrshort}{\ensuremath{1.47}}                        
\newcommand{\hatcurPPrlong}{\ensuremath{1.466\pm0.040}}                
\newcommand{\hatcurPPmrcorr}{\ensuremath{0.08}}                        
\newcommand{\hatcurPPteff}{\ensuremath{1840\pm21}}                     
\newcommand{\hatcurPPtheta}{\ensuremath{0.096\pm0.011}}                
\newcommand{\hatcurPPfluxavg}{\ensuremath{2.59\pm0.12}}                
\newcommand{\hatcurXAv}{\ensuremath{0.0080_{-0.0080}^{+0.0590}}}       
\newcommand{\hatcurXdistred}{\ensuremath{310.5\pm7.1}}                 
\newcommand{\hatcurSMEiteffeccen}{\ensuremath{6516\pm50}}              
\newcommand{\hatcurSMEizfeheccen}{\ensuremath{-0.071\pm0.080}}         
\newcommand{\hatcurSMEizfehshorteccen}{\ensuremath{-0.07}}             
\newcommand{\hatcurSMEiloggeccen}{\ensuremath{4.16\pm0.10}}            
\newcommand{\hatcurSMEivsineccen}{\ensuremath{40.14\pm0.50}}           
\newcommand{\hatcurSMEivmaceccen}{\ensuremath{0.0}}                    
\newcommand{\hatcurSMEivmiceccen}{\ensuremath{0.0}}                    
\newcommand{\hatcurRVecceneccen}{\ensuremath{0.130\pm0.058}}           
\newcommand{\hatcurRVeccentwosiglimeccen}{\ensuremath{<0.246}}         
\newcommand{\hatcur}{HAT-P-56}
\newcommand{\hatcurb}{HAT-P-56b}

\newcommand{\hatcurCCepic}{202126852}
\newcommand{\hatcurCCkepmag}{\ensuremath{10.9}}

\newcommand{\hatcurlumind}{\rhostar}
\newcommand{\hatcurjhkfilset}{ESO}

\newcommand{\hatcurSMEversion}{ii}                                       
\newcommand{\hatcurSMEteff}{\ifthenelse{\equal{\hatcurSMEversion}{i}}{\hatcurSMEiteff}{\hatcurSMEiiteff}}
\newcommand{\hatcurSMEzfeh}{\ifthenelse{\equal{\hatcurSMEversion}{i}}{\hatcurSMEizfeh}{\hatcurSMEiizfeh}}
\newcommand{\hatcurSMEzfehshort}{\ifthenelse{\equal{\hatcurSMEversion}{i}}{\hatcurSMEizfehshort}{\hatcurSMEiizfehshort}}
\newcommand{\hatcurSMElogg}{\ifthenelse{\equal{\hatcurSMEversion}{i}}{\hatcurSMEilogg}{\hatcurSMEiilogg}}
\newcommand{\hatcurSMEvsin}{\ifthenelse{\equal{\hatcurSMEversion}{i}}{\hatcurSMEivsin}{\hatcurSMEiivsin}}
\newcommand{\hatcurSMEvmac}{\ifthenelse{\equal{\hatcurSMEversion}{i}}{\hatcurSMEivmac}{\hatcurSMEiivmac}}
\newcommand{\hatcurSMEvmic}{\ifthenelse{\equal{\hatcurSMEversion}{i}}{\hatcurSMEivmic}{\hatcurSMEiivmic}}

\newcommand{\hatcurSMEteffcirc}{\ifthenelse{\equal{\hatcurSMEversion}{i}}{\hatcurSMEiteffcirc}{\hatcurSMEiiteffcirc}}
\newcommand{\hatcurSMEzfehcirc}{\ifthenelse{\equal{\hatcurSMEversion}{i}}{\hatcurSMEizfehcirc}{\hatcurSMEiizfehcirc}}
\newcommand{\hatcurSMEzfehshortcirc}{\ifthenelse{\equal{\hatcurSMEversion}{i}}{\hatcurSMEizfehshortcirc}{\hatcurSMEiizfehshortcirc}}
\newcommand{\hatcurSMEloggcirc}{\ifthenelse{\equal{\hatcurSMEversion}{i}}{\hatcurSMEiloggcirc}{\hatcurSMEiiloggcirc}}
\newcommand{\hatcurSMEvsincirc}{\ifthenelse{\equal{\hatcurSMEversion}{i}}{\hatcurSMEivsincirc}{\hatcurSMEiivsincirc}}
\newcommand{\hatcurSMEvmaccirc}{\ifthenelse{\equal{\hatcurSMEversion}{i}}{\hatcurSMEivmaccirc}{\hatcurSMEiivmaccirc}}
\newcommand{\hatcurSMEvmiccirc}{\ifthenelse{\equal{\hatcurSMEversion}{i}}{\hatcurSMEivmiccirc}{\hatcurSMEiivmiccirc}}

\newcommand{\hatcurSMEteffeccen}{\ifthenelse{\equal{\hatcurSMEversion}{i}}{\hatcurSMEiteffeccen}{\hatcurSMEiiteffeccen}}
\newcommand{\hatcurSMEzfeheccen}{\ifthenelse{\equal{\hatcurSMEversion}{i}}{\hatcurSMEizfeheccen}{\hatcurSMEiizfeheccen}}
\newcommand{\hatcurSMEzfehshorteccen}{\ifthenelse{\equal{\hatcurSMEversion}{i}}{\hatcurSMEizfehshorteccen}{\hatcurSMEiizfehshorteccen}}
\newcommand{\hatcurSMEloggeccen}{\ifthenelse{\equal{\hatcurSMEversion}{i}}{\hatcurSMEiloggeccen}{\hatcurSMEiiloggeccen}}
\newcommand{\hatcurSMEvsineccen}{\ifthenelse{\equal{\hatcurSMEversion}{i}}{\hatcurSMEivsineccen}{\hatcurSMEiivsineccen}}
\newcommand{\hatcurSMEvmaceccen}{\ifthenelse{\equal{\hatcurSMEversion}{i}}{\hatcurSMEivmaceccen}{\hatcurSMEiivmaceccen}}
\newcommand{\hatcurSMEvmiceccen}{\ifthenelse{\equal{\hatcurSMEversion}{i}}{\hatcurSMEivmiceccen}{\hatcurSMEiivmiceccen}}

\newcounter{planetcounter}

\newboolean{emulateapj}
\setboolean{emulateapj}{true}

\newboolean{rvtablelong}
\setboolean{rvtablelong}{true}

\newboolean{astroph}
\setboolean{astroph}{true}

\shortauthors{Huang et al.}
\shorttitle{
\hatcur\lowercase{b}
}
\ifthenelse{\boolean{emulateapj}}{
    \newcommand{\titledag}{$\dagger$}
}{
    \newcommand{\titledag}{\dagger}
}

\begin{document}
\title{
\hatcur\lowercase{b}: 
An inflated massive Hot Jupiter transiting a bright F
star followed up with {\em K2} Campaign 0 observations

\altaffilmark{\titledag} 
}
\author{
    C.~X.~Huang\altaffilmark{1},
    J.~D.~Hartman\altaffilmark{1},
    G.~\'A.~Bakos\altaffilmark{1,6,7},
    K.~Penev\altaffilmark{1},
    W.~Bhatti\altaffilmark{1},
    A.~Bieryla\altaffilmark{2},
    M.~de Val-Borro\altaffilmark{1},
    D.~W.~Latham\altaffilmark{2},
    L.~A.~Buchhave\altaffilmark{2,3},
    Z.~Csubry\altaffilmark{1},
    G.~Kov\'acs\altaffilmark{4},
    B.~B\'eky\altaffilmark{2},
    E.~Falco\altaffilmark{2},
    P.~Berlind\altaffilmark{2},
    M.~L.~Calkins\altaffilmark{2},
    G.~A.~Esquerdo\altaffilmark{2},
    J.~L\'az\'ar\altaffilmark{5},
    I.~Papp\altaffilmark{5},
    P.~S\'ari\altaffilmark{5}
}

\altaffiltext{1}{Department of Astrophysical Sciences, Princeton
  University, Princeton, NJ 08544; email: chelsea@astro.princeton.edu}
\altaffiltext{2}{Harvard-Smithsonian Center for Astrophysics, Cambridge,
MA 02138 USA; email: abieryla@cfa.harvard.edu} 
\altaffiltext{3}{Centre for Star and Planet Formation, Natural History 
Museum of Denmark, University of Copenhagen, DK-1350 Copenhagen,
Denmark}
\altaffiltext{4}{Konkoly Observatory, Budapest, Hungary} 
\altaffiltext{5}{Hungarian Astronomical Association (HAA)}
\altaffiltext{6}{Sloan Fellow}
\altaffiltext{7}{Packard Fellow}

\altaffiltext{$\dagger$}{
Based on observations obtained with the Hungarian-made Automated
Telescope Network. Based in part on observations obtained with
the Tillinghast Reflector 1.5\,m telescope and the 1.2\,m telescope,
both operated by the Smithsonian Astrophysical Observatory at the Fred
Lawrence Whipple Observatory in Arizona. Based in part on observations 
obtained with the Apache Point Observatory 3.5-meter telescope, 
which is owned and operated by the Astrophysical Research Consortium. 
Based in part on observations obtained with the {\em Kepler} Space 
Craft in the {\em K2} Campaign 0 Mission.
}


\begin{abstract}

\setcounter{footnote}{10}
We report the discovery of \hatcurb\ by the HATNet survey, an inflated
hot Jupiter transiting a bright F type star in Field 0 of NASA's {\em
K2} mission.  We combine ground-based discovery and follow-up \lcs\
with high precision photometry from {\em K2}, as well as ground-based
radial velocities from TRES on the FLWO~1.5\,m telescope to determine the
physical properties of this system.  
\hatcurb{} has a mass of \hatcurPPmshort\,\mjup, radius of 
\hatcurPPrshort\,\rjup, and transits its host star on a near-grazing
orbit with a period of \hatcurLCPshort\,d.  The radius of
\hatcurb{} is among the largest known for a planet with $M_p >
2$\,\mjup.  The host star has a $V$-band magnitude of
\hatcurCCtassmvshort, mass of \hatcurISOmshort\,\msun, and radius of
\hatcurISOrshort\,\rsun.  The periodogram of the {\em K2} \lc\ suggests
the star is a $\gamma$ Dor variable.  \hatcurb\ is an example of a
ground-based discovery of a transiting planet, where space-based
observations greatly improve the confidence in the confirmation of its
planetary nature, and also improve the accuracy of the planetary
parameters.
\setcounter{footnote}{0}
\end{abstract}



\section{Introduction}
\label{sec:introduction}

Searching for transits is one of the most productive methods of
detecting planets outside of our solar system.  Transiting exoplanets
(TEPs) around bright stars are of particular interest as they are the
best targets for follow-up observations to investigate their detailed
orbital geometries, atmospheric properties and chemical compositions.

Recent years have seen a rapid increase in the discovery rate of transit
exoplanets (TEPs).  The {\em Kepler Mission} \citep{borucki:2009} 
has identified $\sim$4000 planetary candidates in its four 
years of operations \citep{mullally:2015}. Approximately a 
hundred of the {\em Kepler} planets have their mass measured using 
stellar radial velocities 
(\citet{Batalha:2011}, \citet{Gautier:2012}, \citet{Marcy:2014}, 
and many others), or numerically modeled transit timing variations 
(\citet{Lissauer:2011}, \citet{Carter:2012}, \citet{Wu:2013}, 
and many others). 
The majority of {\em Kepler} 
planet candidate hosting stars are, however, often faint, 
complicating both the accurate planetary
mass measurements and also the study of their physical nature. For 
example, to date, the majority of the planets with precise mass
measurements are still from the ground-based wide field transiting
planet surveys such as SuperWASP \citep{pollacco:2006}, 
HATNet/HATSouth \citep{bakos:2004:hatnet,bakos:2013:hatsouth}, 
KELT \citep{pepper:2007}, and
others.

Traditionally, confirmation of planets from ground-based transit
surveys relies on high precision photometric follow-up from one or two
meter class ground-based telescopes.  These follow-up observations
ensure the robustness of the detections, help to rule out false
positives due to binaries, and further constrain the planet parameters. 
However, due to the constraints of ground-based observations, it's
usually difficult to observe the full orbital phase of a transiting
planet candidate, including primary and secondary transits.  The
presence or lack of deep secondary eclipses, or out-of-transit
variations is often important information for ruling out blended
stellar eclipsing binary scenarios.  The quality of ground-based
photometry may also be reduced due to poor weather conditions, site and
target restrictions, and airmass trends.

Due to the loss of two reaction wheels during the main mission, the
{\em Kepler} spacecraft entered a new observation phase.  This
successor mission of {\em Kepler}, called the {\em K2} mission
\citep{howell:2014}, covers a much larger area of sky with a step and
stare strategy.  The Guest Observation mode of this mission provides a
great tool to follow up the ground-based planetary candidates, and
enables a new method of discovering TEPs through a synergy between
ground- and space-based transit surveys.  {\em K2} observations allow
us to obtain high precision \lcs\ of candidates over a continuous and
relatively long time-baseline covering many transit events due 
to a planet.

Once the planet is confirmed, having high precision \lcs\ from space
observations will also enable further characterization of the planet.  
Confirmed
planets from the ground have been recognized as valuable targets to be
followed up by {\em K2} \citep{bakos:2014:hat54,brown:2014}.  The
planets discovered by ground based TEP surveys, and also observed by
the original {\em Kepler Mission}, such as TrEs-2b 
\citep{odonovan:2006},  HAT-P-7b \citep{pal:2008:hat7}, and 
HAT-P-11b \citep{bakos:2010:hat11}, are among the best studied planets.  
With the high precision {\em Kepler} light curves, many effects due to the
planets can be measured, such as the orbital phase variation and
occultation of the planet \citep{barclay:2012,jackson:2012,
kipping:2011,welsh:2010,borucki:2009}.  The spin-orbit obliquity angle
of HAT-P-11b has also been constrained using the starspot crossing
events observed in the {\em Kepler} light curves
\citep{sanchisojeda:2011,deming:2011}.  Detailed modeling of the
transit shape using the Short Cadence photometry of {\em Kepler} can
also be used to determine the probable gravity darkening effect, and
constrain the oblateness of the planet
\citep{morris:2013,masuda:2015,zhu:2014}.

Following up candidates from ground-based transit surveys is an
efficient way to utilize the {\em K2} observational resources.  Due to
constraints from the new mode of operation, the number of targets
observed by {\em K2} per field is much smaller than for the original
{\em Kepler} field.  The high priority planetary candidates from
ground-based surveys are pre-selected targets which are known to show
transits, and which have already been vetted against various false positive
scenarios.  Selecting these candidates to fall on {\em K2} 
``postage stamps" is
one way to increase the confirmed planet yield from this mission.

\begin{figure}[]
\plotone{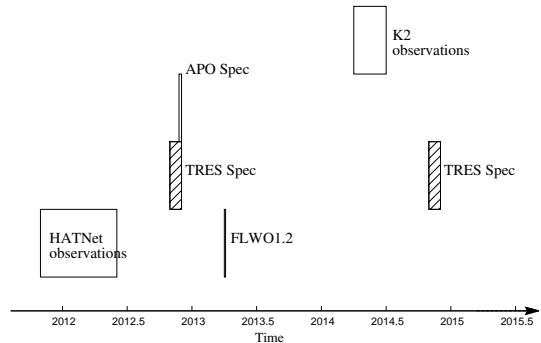}
\caption[]{
   The sequence of observations that lead to the discovery and 
   confirmation flow of \hatcurb{}. 
\label{fig:flowchart}}
\end{figure}

In this paper we present the discovery (see Figure \ref{fig:flowchart}) of a transiting planet,
\hatcurb, in the {\em K2} Campaign 0 field.  This planet was originally
identified as a HATNet \citep{bakos:2004:hatnet} planetary candidate,
was followed up by the TRES spectrograph on the FLWO~1.5\,m telescope,
and also by the KeplerCam imager on the FLWO~1.2\,m telescope. 
Encouraged by these initial results, all pointing toward a {\em bona
fide} planet orbiting the host star \hatcur, the target was proposed
for {\em K2} observations.  Indeed, the very high quality photometric
observations of {\em K2} confirmed the transit, and also eliminated
most of the possible blend scenarios.  We then continued following-up
\hatcur\ with the TRES spectrograph, so as to determine the mass of the
orbiting body.

In \S \ref{sec:obs} we summarize the detection of the photometric
transit signal in the HATNet \lc, follow-up photometry from the ground
and from {\em K2} campaign 0, and our spectroscopic follow-up. 
Analyses of the results are presented in \S \ref{sec:analysis}.  We
show in \S\ref{sec:discussion} that \hatcurb{} is one of the most
inflated objects observed that belong to the massive hot Jupiter
population ($M>2M_J$), even when we take into account of the 
amount of irradiation from the host star. 


\section{Observations}
\label{sec:obs}

\subsection{Ground Based Photometry}
\label{sec:photometry}

All time-series photometric data that we collected for \hatcur{} are
provided in Table~\ref{tab:phfu}. We discuss these observations below.
All of our discovery data and follow-up data are publicly available in
electronic format as machine readable tables, and  also via HATNet website\footnote{\url{http://hatnet.org}}.

\subsubsection{Photometric detection}
\label{sec:detection}

The star \hatcur{} was observed by the HATNet wide-field photometric
instruments \citep{bakos:2004:hatnet} between the nights of UT 2011
October 14 and UT 2012 May 3. A total of 6509 observations of a
$10\fdg 6 \times 10\fdg 6$ field centered at ${\rm R.A.} = 06^{\rm
  hr}24^{\rm min}$, ${\rm Dec.} = +30^{\circ}$ were made with the
HAT-6 telescope in Arizona, and 4194 observations of this same field
were made with the HAT-9 telescope in Hawaii (the count is after
filtering some $40$ outlier measurements). We used a Sloan~$r$ filter
and an exposure time of $180$\,s. Following \citet{bakos:2010:hat11}
and \citet{kovacs:2005:TFA}, we reduced the images to trend-filtered
light curves for the $\sim$ 124,000 stars in the field 
with $r < 14.5$\,mag,
achieving a point-to-point r.m.s.~precision of 3.5\,mmag for the
brightest non-saturated stars with $r \sim 10$\,mag (for \hatcur{} the
r.m.s.~of the residuals from our best-fit transit model is
6.7\,mmag). We searched these light curves for periodic transit
signals using the Box Least Squares algorithm
\citep[BLS;][]{kovacs:2002:BLS}.

\begin{figure}[]
\plotone{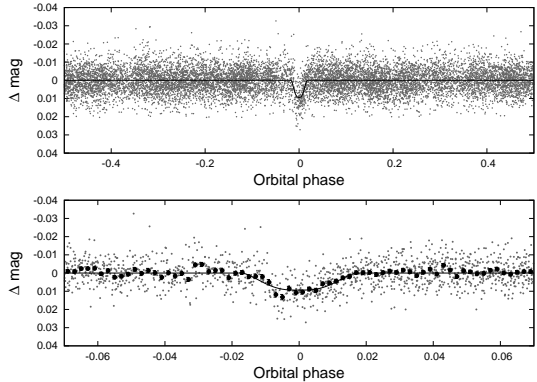}
\caption[]{
    HATNet \lc{} of \hatcur\ phase folded with the transit period.
    The top panel shows the unbinned light curve,
    while the bottom shows the region zoomed-in on the transit, with
    dark filled circles for the light curve binned in phase with a
    binsize of 0.002. The solid line shows the model fit to the light
    curve.
\label{fig:hatnet}}
\end{figure}

A total of 29 candidate transiting planets were identified from these
light curves, including \hatcur{}, whose phase-folded HATNet light
curve we show in Figure~\ref{fig:hatnet}. Reconnaissance spectroscopy
and photometric follow-up observations have been carried out for most
of these candidates, based on which we have rejected 21 of them as
false positives. In addition to \hatcur{}, which we confirm as a
planetary system in this paper, one other planet has been confirmed
from this field \citep[HAT-P-54b;][]{bakos:2014:hat54}. Six of the
candidates remain active.

\subsubsection{Photometric follow-up with KeplerCam}
\label{sec:phfu}

\begin{figure}[]
\plotone{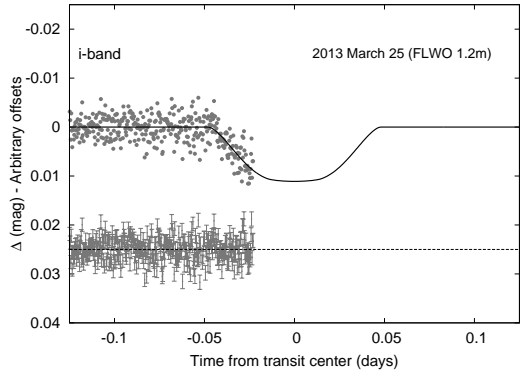}
\caption{
    Unbinned transit light curve for \hatcur, acquired with KeplerCam
    at the \flwof{} telescope on the night of 2013 March 25.  The light
    curve was corrected for trends, which were fit simultaneously with
    the transit model.  Our best fit from the global modeling
    described in \refsecl{analysis} is shown by the solid line. 
    Residuals from the fit are displayed below the original light
    curve.  The error bars represent the photon and background shot
    noise, plus the readout noise.
}
\label{fig:lc}
\end{figure}

Initial photometric follow-up observations of \hatcur{} were carried
out with KeplerCam on the Fred Lawrence Whipple Observatory (FLWO)
1.2\,m telescope.  We observed a single transit ingress on the night of
UT 2013 March 25 using an $i$-band filter and an exposure time of $10$\,s. 
The images were reduced to a light curve following
\citet{bakos:2010:hat11}, including external parameter decorrelation
performed simultaneously with the transit fit to remove systematic
trends; this trend-corrected light curve is shown in Figure~\ref{fig:lc}. 
The r.m.s.~of the residuals from our best-fit model is 2.2\,mmag for
these data.

\subsection{Spectroscopy}
\label{sec:hispec}

We carried out spectroscopic observations of \hatcur{} between UT 2012
October 31 and UT 2014 November 25 using the Tillinghast Reflector
Echelle Spectrograph \citep[TRES;][]{furesz:2008} on the 1.5\,m
Tillinghast Reflector at FLWO.  We extracted spectra from the images
and measured initial RVs, bisector spans (BSs) and stellar atmospheric
parameters following \cite{buchhave:2010:hat16}.

The first three TRES observations made between UT 2012 October 31 and
UT 2012 November 7 using a short exposure time of 360\,s resulted in a
S/N per resolution element of 38--45, and were used for reconnaissance
purposes.  Based on these three reconnaissance spectra we found that
\hatcur{} is an F dwarf star with a fairly rapid projected rotation
velocity of $\sim 40$\,\kms, and no evidence for additional stellar
components in the spectrum.  The three RV measurements were consistent
with no variation above 600\,\ms\ (3$\sigma$ upper limit).  A single
APO~3.5\,m/ARCES spectrum of \hatcur{} was also obtained for
reconnaissance on UT 2012 November 7.  A 160\,s exposure time was used
to achieve a S/N per resolution element of 28.6.  We reduced the
observation to a wavelength-calibrated spectrum with the IRAF {\sc
echelle} package\footnote{
IRAF is distributed by the National Optical Astronomy Observatories,
which are operated by the Association of Universities for Research in
Astronomy, Inc., under cooperative agreement with the National Science
Foundation.
} and used the Stellar Parameter Classification program
\citep[SPC;][]{buchhave:2012:spc} to measure the atmospheric
parameters and radial velocity from the spectrum. This spectrum was
also found to be single-lined, had an RV consistent with the TRES
measurements, and also indicated that the target was a F dwarf with 
$\vsini = 40$\,\kms.

\begin{figure} []
\plotone{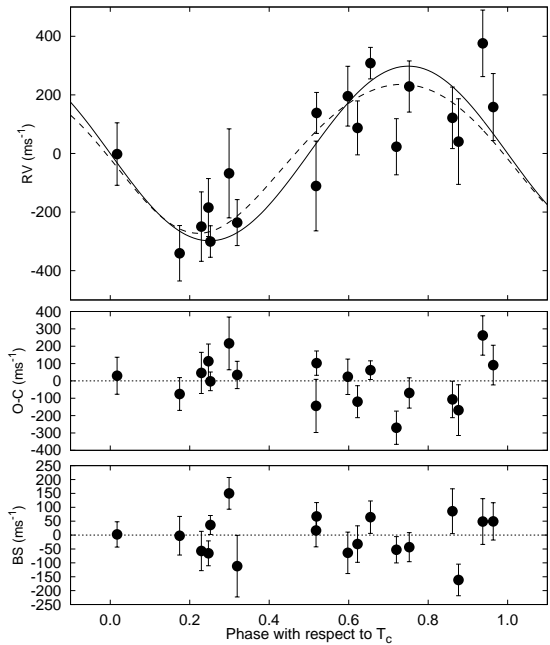}
\caption{
    {\em Top panel:} RV measurements from FLWO~1.5\,m/TRES shown as a
    function of orbital phase, along with our best-fit circular model
    (solid line; see \reftabl{planetparam}), and our best-fit
    eccentric model (dashed line).  Zero phase corresponds to the time
    of mid-transit.  The center-of-mass velocity has been subtracted.
    {\em Second panel:} Velocity $O\!-\!C$ residuals from the best
    fit circular orbit. 
    {\em Third panel:} Bisector spans (BS).
    Note the different vertical scales of the panels.
}
\label{fig:rvbis}
\end{figure}

Following the {\em K2} observations (Section~\ref{sec:kepphfu}), 
which showed clean transits with no evidence 
of secondary eclipses or strong out-of-transit variability in phase 
with the transits, we resumed spectroscopic monitoring with 
TRES on the FLWO~1.5\,m telescope, now with the aim of
measuring the mass of the planet by detecting the orbital motion of the
host star.  A total of 18 high S/N (ranging from 50--113) observations
were collected between UT 2014 October 2 and UT 2014 November 15.  We
measured stellar atmospheric parameters from these spectra using SPC, 
and carried out a multi-order velocity analysis following
\citet{bieryla:2014:hat49} to measure the RVs relative to one of 
the observed spectra and spectral line bisector spans (BSs). 
Table~\ref{tab:rvs} provides the measurements
extracted from these spectra, where we exclude the reconnaissance
spectra that were deemed to be of too low S/N to be used for our final
characterization of the orbit and the stellar atmospheric properties. 
The phase-folded RVs and BSs are shown in Figure~\ref{fig:rvbis}
together with our best-fit circular and eccentric orbit models. 
We show in Section~\ref{sec:globmod} that the above RV measurements 
confirm the planet nature of \hatcurb.

\setcounter{planetcounter}{1}
%

\ifthenelse{\boolean{emulateapj}}{
    \begin{deluxetable*}{lrrrrr}
}{
    \begin{deluxetable}{lrrrrr}
}
\tablewidth{0pc}
\tablecaption{
    Differential photometry of
    \hatcur\label{tab:phfu}.
}
\tablehead{
    \colhead{BJD\tablenotemark{a}} & 
    \colhead{Phot\tablenotemark{b}} & 
    \colhead{\ensuremath{\sigma_{\rm Phot}}} &
    \colhead{Phot(orig)\tablenotemark{c}} & 
    \colhead{Filter} &
    \colhead{Instrument} \\
    \colhead{\hbox{~~~~(2,400,000$+$)~~~~}} & 
    \colhead{} & 
    \colhead{} &
    \colhead{} & 
    \colhead{} & 
    \colhead{}
}
\startdata
$ 55954.98325 $ & $   0.00245 $ & $   0.00430 $ & $ \cdots $ & $  r$ &     HATNet\\
$ 55971.72852 $ & $   0.00273 $ & $   0.00463 $ & $ \cdots $ & $  r$ &     HATNet\\
$ 55860.09538 $ & $   0.00383 $ & $   0.00396 $ & $ \cdots $ & $  r$ &     HATNet\\
$ 55957.77449 $ & $   0.00368 $ & $   0.00464 $ & $ \cdots $ & $  r$ &     HATNet\\
$ 55929.86619 $ & $   0.00725 $ & $   0.00426 $ & $ \cdots $ & $  r$ &     HATNet\\
$ 55901.95813 $ & $  -0.00271 $ & $   0.00432 $ & $ \cdots $ & $  r$ &     HATNet\\
$ 55888.00414 $ & $  -0.01589 $ & $   0.00408 $ & $ \cdots $ & $  r$ &     HATNet\\
$ 55904.75041 $ & $   0.00110 $ & $   0.00603 $ & $ \cdots $ & $  r$ &     HATNet\\
$ 55929.86812 $ & $   0.00151 $ & $   0.00373 $ & $ \cdots $ & $  r$ &     HATNet\\
$ 55901.95982 $ & $   0.00766 $ & $   0.00400 $ & $ \cdots $ & $  r$ &     HATNet\\
\enddata
\tablenotetext{a}{
    Barycentric Julian Date calculated directly from UTC, {\em
      without} correction for leap seconds for HATNet and KeplerCam.
}
\tablenotetext{b}{
    For HATNet and KeplerCam this is in units of magnitudes, for {\em
      K2} it is in relative flux. The out-of-transit level has been
    subtracted. These values have been corrected for trends
    simultaneously with the transit fit for the follow-up data. For
    HATNet trends were filtered {\em before} fitting for the transit.
}
\tablenotetext{c}{
    Raw photometry values after correction using comparison stars, but
    without additional trend-filtering. For KeplerCam this is in
    magnitudes, for {\em K2} it is in relative flux. We do not report this
    value for HATNet.
}
\tablecomments{
    This table is available in a machine-readable form in the online
    journal.  A portion is shown here for guidance regarding its form
    and content. Data is also available at \url{hatnet.org}.
}
\ifthenelse{\boolean{emulateapj}}{
    \end{deluxetable*}
}{
    \end{deluxetable}
}

\ifthenelse{\boolean{emulateapj}}{
    \begin{deluxetable*}{lrrrrrr}
}{
    \begin{deluxetable}{lrrrrrr}
}
\tablewidth{0pc}
\tablecaption{
    Relative radial velocities, and bisector span measurements of \hatcur.
    \label{tab:rvs}
}
\tablehead{
    \colhead{BJD\tablenotemark{a}} &
    \colhead{RV\tablenotemark{b}} &
    \colhead{\ensuremath{\sigma_{\rm RV}}} &
    \colhead{BS} &
    \colhead{\ensuremath{\sigma_{\rm BS}}} &
    \colhead{Phase} &
    \colhead{Instrument}\\
    \colhead{\hbox{(2,456,900$+$)}} &
    \colhead{(\ms)} &
    \colhead{(\ms)} &
    \colhead{(\ms)} &
    \colhead{(\ms)} &
    \colhead{} &
    \colhead{}
}
\startdata
$ 32.99501 $ & $   68 $ & $   114 $ & $    48.70 $ & $    82.40 $ & $   0.937 $ & TRES \\
$ 34.00383 $ & $   -376 $ & $   152 $ & $   150.10 $ & $    57.10 $ & $   0.299 $ & TRES \\
$ 34.99731 $ & $   0 $ & $    54 $ & $    64.40 $ & $    58.30 $ & $   0.655 $ & TRES \\
$ 36.00811 $ & $    -310 $ & $   107 $ & $     2.40 $ & $    45.50 $ & $   0.017 $ & TRES \\
$ 42.99104 $ & $   -170 $ & $    70 $ & $    67.10 $ & $    50.00 $ & $   0.519 $ & TRES \\
$ 43.98771 $ & $    -268 $ & $   146 $ & $  -161.50 $ & $    56.80 $ & $   0.876 $ & TRES \\
$ 44.97264 $ & $  -558 $ & $   119 $ & $   -57.30 $ & $    70.80 $ & $   0.229 $ & TRES \\
$ 46.00112 $ & $   -113 $ & $   102 $ & $   -63.80 $ & $    74.60 $ & $   0.598 $ & TRES \\
$ 58.97630 $ & $  -493 $ & $    99 $ & $   -65.60 $ & $    44.80 $ & $   0.247 $ & TRES \\
$ 60.02365 $ & $    -221 $ & $    92 $ & $   -32.30 $ & $    65.70 $ & $   0.622 $ & TRES \\
$ 60.97679 $ & $   -150 $ & $   114 $ & $    49.20 $ & $    67.20 $ & $   0.964 $ & TRES \\
$ 61.96910 $ & $  -544 $ & $    78 $ & $  -111.70 $ & $   111.00 $ & $   0.319 $ & TRES \\
$ 65.96895 $ & $   -80 $ & $    87 $ & $   -43.50 $ & $    52.20 $ & $   0.753 $ & TRES \\
$ 69.93743 $ & $  -649 $ & $    95 $ & $    -2.40 $ & $    69.50 $ & $   0.174 $ & TRES \\
$ 70.89553 $ & $  -419 $ & $   153 $ & $    16.40 $ & $    58.90 $ & $   0.518 $ & TRES \\
$ 71.85396 $ & $   -187 $ & $   105 $ & $    85.90 $ & $    80.70 $ & $   0.861 $ & TRES \\
$ 72.94426 $ & $  -609 $ & $    54 $ & $    36.40 $ & $    34.10 $ & $   0.252 $ & TRES \\
$ 77.04183 $ & $   -331 $ & $    96 $ & $   -52.80 $ & $    47.40 $ & $   0.720 $ & TRES \\
\enddata
\tablenotetext{a}{
    Barycentric Julian Date calculated directly from UTC, {\em
      without} correction for leap seconds.
}
\tablenotetext{b}{
    The zero-point of these velocities is arbitrary. An overall offset
    $\gamma_{\rm rel}$ fitted to these velocities in \refsecl{analysis}
    has {\em not} been subtracted. 
}
\ifthenelse{\boolean{rvtablelong}}{
}{
} 
\ifthenelse{\boolean{emulateapj}}{
    \end{deluxetable*}
}{
    \end{deluxetable}
}

\ifthenelse{\boolean{emulateapj}}{
    \begin{deluxetable*}{lrrrrrrrr}
}{
    \begin{deluxetable}{lrrrrrrrr}
}
\tablewidth{0pc}
\tablecaption{
    Stellar Atmospheric Parameters for \hatcur{} Measured with SPC.
    \label{tab:spc}
}
\tablehead{
    \colhead{BJD\tablenotemark{a}} &
    \colhead{T$_{\rm eff,1}$ \tablenotemark{b}} &
    \colhead{$\log g_{1}$ \tablenotemark{b}} &
    \colhead{[m/H]$_{1}$ \tablenotemark{b}} &
    \colhead{$v \sin i_{\rm eff,1}$ \tablenotemark{b}} &
    \colhead{T$_{\rm eff,2}$ \tablenotemark{c}} &
    \colhead{[m/H]$_{2}$ \tablenotemark{c}} &
    \colhead{$v \sin i_{\rm eff,2}$ \tablenotemark{c}} &
    \colhead{S/N \tablenotemark{c}} \\
    \colhead{\hbox{(2,456,900$+$)}} &
    \colhead{(K)} &
    \colhead{(c.g.s.)} &
    \colhead{} &
    \colhead{(\kms)} &
    \colhead{(K)} &
    \colhead{} &
    \colhead{(\kms)} &
    \colhead{} \\
}
\startdata
$ 32.99501 $ & $6406$ & $4.01$ & $-0.140$ & $40.06$ & $6545$ & $-0.052$ & $39.84$ & $55.7$ \\
$ 34.00383 $ & $6500$ & $4.09$ & $-0.029$ & $40.72$ & $6592$ & $+0.034$ & $40.67$ & $56.2$ \\
$ 34.99731 $ & $6500$ & $4.15$ & $-0.174$ & $40.19$ & $6538$ & $-0.150$ & $40.10$ & $102.7$ \\
$ 36.00811 $ & $6588$ & $4.27$ & $-0.103$ & $39.83$ & $6576$ & $-0.116$ & $39.87$ & $84.0$ \\
$ 42.99104 $ & $6528$ & $4.17$ & $-0.100$ & $39.62$ & $6572$ & $-0.079$ & $39.54$ & $92.9$ \\ 
$ 43.98771 $ & $6527$ & $4.21$ & $-0.101$ & $40.20$ & $6553$ & $-0.099$ & $40.18$ & $83.8$ \\ 
$ 44.97264 $ & $6533$ & $4.23$ & $-0.104$ & $39.86$ & $6531$ & $-0.100$ & $39.89$ & $80.1$ \\ 
$ 46.00112 $ & $6551$ & $4.21$ & $-0.095$ & $39.75$ & $6569$ & $-0.091$ & $39.74$ & $101.6$ \\
$ 58.97630 $\tablenotemark{d} & $\cdots$ & $\cdots$ & $\cdots$ & $\cdots$ & $6555$ & $-0.071$ & $40.38$ & $91.9$ \\  
$ 60.02365 $\tablenotemark{d} & $\cdots$ & $\cdots$ & $\cdots$ & $\cdots$ & $6598$ & $-0.057$ & $40.15$ & $74.9$ \\  
$ 60.97679 $\tablenotemark{d} & $\cdots$ & $\cdots$ & $\cdots$ & $\cdots$ & $6534$ & $-0.109$ & $39.59$ & $90.6$ \\  
$ 61.96910 $\tablenotemark{d} & $\cdots$ & $\cdots$ & $\cdots$ & $\cdots$ & $6544$ & $+0.062$ & $41.80$ & $49.3$ \\   
$ 65.96895 $\tablenotemark{d} & $\cdots$ & $\cdots$ & $\cdots$ & $\cdots$ & $6576$ & $-0.082$ & $40.03$ & $113.2$ \\ 
$ 69.93743 $\tablenotemark{d} & $\cdots$ & $\cdots$ & $\cdots$ & $\cdots$ & $6603$ & $-0.077$ & $40.02$ & $91.9$ \\  
$ 70.89553 $\tablenotemark{d} & $\cdots$ & $\cdots$ & $\cdots$ & $\cdots$ & $6612$ & $-0.083$ & $39.80$ & $78.6$ \\  
$ 71.85396 $\tablenotemark{d} & $\cdots$ & $\cdots$ & $\cdots$ & $\cdots$ & $6540$ & $-0.093$ & $39.88$ & $97.6$ \\  
$ 72.94426 $\tablenotemark{d} & $\cdots$ & $\cdots$ & $\cdots$ & $\cdots$ & $6592$ & $-0.123$ & $39.59$ & $96.5$ \\  
$ 77.04183 $\tablenotemark{d} & $\cdots$ & $\cdots$ & $\cdots$ & $\cdots$ & $6561$ & $-0.099$ & $40.03$ & $78.4$ \\  
\enddata
\tablenotetext{a}{
    Barycentric Julian Date calculated directly from UTC, {\em
      without} correction for leap seconds.
}
\tablenotetext{b}{
    Measurements from initial SPC iteration in which $\loggstar$ was
    allowed to vary.
}
\tablenotetext{c}{
    Measurements from second SPC iteration in which we fixed $\loggstar$
    to 4.235.
}
\tablenotetext{d}{
    Observations were obtained after running our initial SPC
    analysis. They were included, however, in our fixed-$\loggstar$ SPC
    analysis.
}
\ifthenelse{\boolean{rvtablelong}}{
}{
} 
\ifthenelse{\boolean{emulateapj}}{
    \end{deluxetable*}
}{
    \end{deluxetable}
}

\subsection{{\em K2} photometry follow up}
\label{sec:kepphfu}
\begin{figure}[]
\plotone{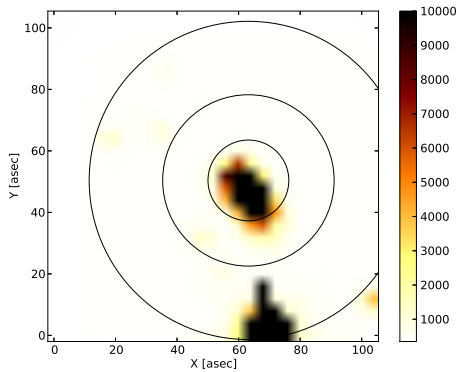}
\caption{
    A image stamp from {\em K2} Campaign 0 containing \hatcur{}. 
    The neighbor at 49\arcsec away from \hatcur{} is partially 
    observed at the bottom of the
    stamp.  The inner most circle indicates the optimal aperture we
    selected that gave the best out of transit light curve r.m.s.. 
    The annulus between the outer two circles is used to 
    calculate the background.
}
\label{fig:kepstamp}
\end{figure}

Encouraged by the HATNet, FLWO 1.2\,m and TRES observations, we
proposed \hatcur{} as a target for the {\em K2} Campaign 0 through the
{\em Kepler} Guest Observing Program.  The observations are in 
{\em Kepler} Long Cadence mode ($\sim$ 30 min exposures) with 
a stamp size of 27$\times$27 pixels on {\em Kepler} CCD 
Module 10, Channel 29, and were carried out between 
BJD 2456728.5282 and BJD 2456805.1883 (UT 2014
March 8 to UT 2014 May 27).  There are two data gaps during the
observation, from BJD 2456732.4309 to BJD 2456735.6386, and from BJD
2456744.1180 to BJD 2456767.5941.  Module 10 is one of the outermost
modules on the {\em Kepler} spacecraft, and as a result, the target
drifted for a significant fraction of a pixel during each 30\,min
exposure, leading to an elongated PSF.  The bright neighbor 49\arcsec\
away from the target is partially observed in the target 
``postage stamps" (see
Figure \ref{fig:kepstamp}).

The data were reduced using HATNet's reduction pipeline, 
certain aspects of which are described in \citet{pal:2009:thesis}.  After
source extraction, we measured the flux of the star in a series of
circular apertures, and the sky background in circular annuli. 
The sky background is determined by taking the 
median of all the pixels with iterative outlier rejection in 
the annuli to exclude the influence 
from the bright neighbor star. 
The best aperture was of 3.3 pixels radius, based on the 
minimum r.m.s.~we achieve (shown in
Figure \ref{fig:kepstamp} together with the background annulus we
used).

To correct the photometry for variations due to the motion of the
spacecraft, we performed an External Parameter Decorrelation (EPD) on
the extracted raw \lc.  Before detrending, data points obtained during
thrust fires were rejected following a similar methodology to that
described in \citet{vanderburg:2014}. 
We also discarded the first segment of data 
(from BJD 2456728.752975 to BJD 2456771.394346) obtained 
before the first safe mode, and the first transit right 
after the second safe mode while the space craft was still 
adjusting its pointing. We describe the systematics as a 
function of the $x$, $y$ centroid of the star, the background flux $bg$, 
and the uncertainty of the background flux $e_{bg}$:
\begin{equation}
\begin{aligned}
f(m) &= c_0+c_1\sin(2\pi\,x)+c_2\cos(2\pi\,x)\nonumber\\
    &+c_3\sin(2\pi\,y)+c_4\cos(2\pi\,y)\nonumber \\
    &+c_5\sin(4\pi\,x)+c_6\cos(4\pi\,x)\nonumber\\
    &+c_7\sin(4\pi\,y)+c_8\cos(4\pi\,y)\nonumber \\
    &+c_9bg+c_{10}e_{bg}.
\end{aligned}
\end{equation}

To preserve the signal of the transit, we modify the original EPD
algorithm to only fit with the out-of-transit part of the \lc. The 
\lc\ is then detrended with cosine filters as described in
\citet{Huang:2013}.  We allowed a minimum period of 0.5 days for the
filters, to preserve the transit signal. 

\begin{figure*}[]
\plotone{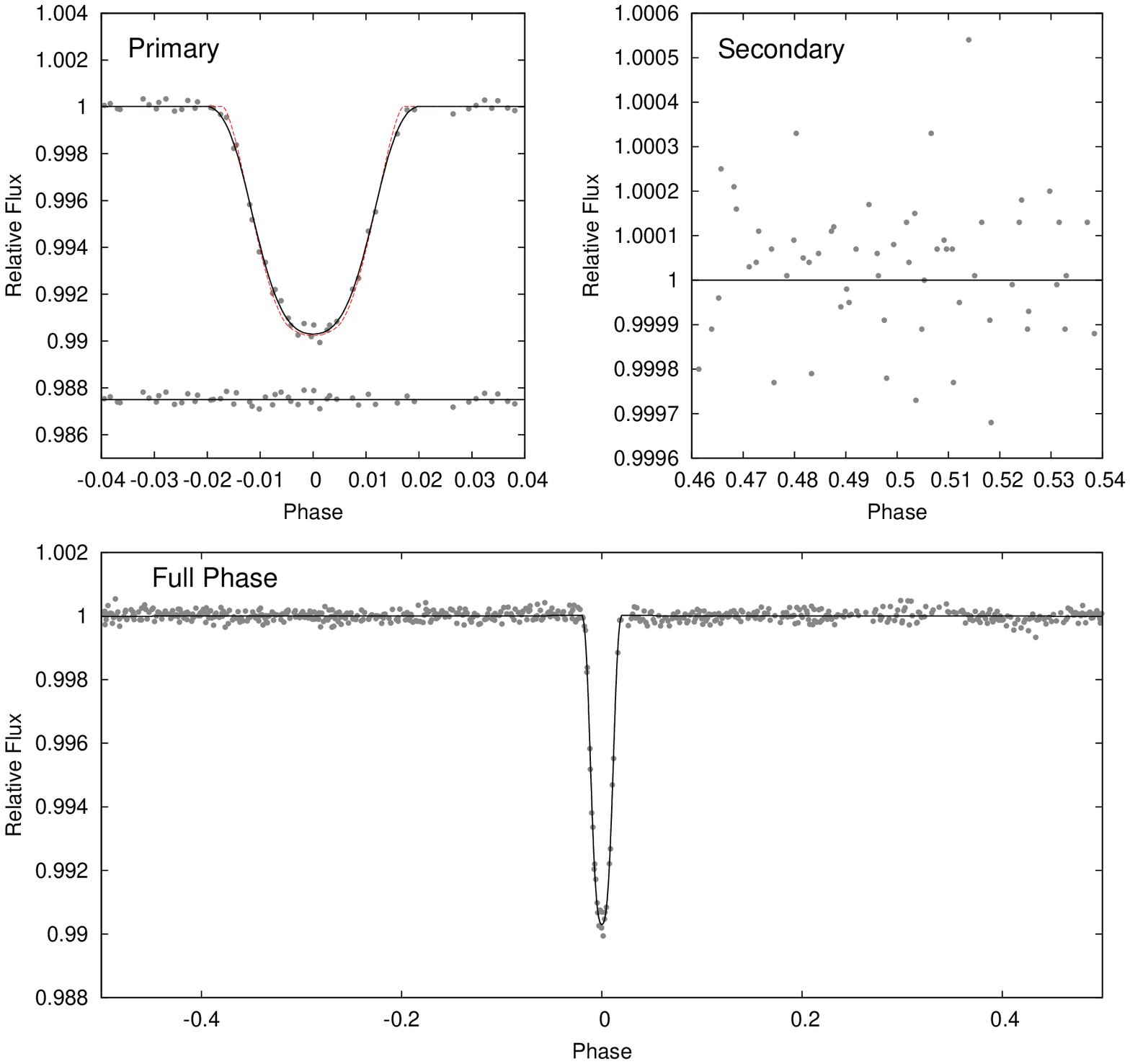}
\caption{
    Phase-folded 30\,min cadence {\em K2} observations of \hatcur{}. 
    The phase-folded light curve is shown in the bottom panel.  
    In the top left we show a zoom-in on the transit, while in the 
    top right we show a
    zoom-in on the expected phase of occultation assuming a circular
    orbit.  The solid line shows our best-fit model to the
    observations, binned over the 30\,min integrations.  The dashed
    line, visible only in the top-left panel, 
    shows the model without binning.
}
\label{fig:keplc}
\end{figure*}

The resulting {\em K2} \lc\ is shown in Figure~\ref{fig:keplc}, 
together with our best-fit transit model.  The residuals have per-point 
r.m.s.~of 0.17\,mmag.  In order to search for additional transit signals, 
we removed the signal of \hatcurb{} from the {\em K2} \lc\
according to the best fit transit parameters, and then reprocessed the
\lc\ with the same detrending steps.  We ran BLS on the detrended
residual \lc.  There is no evidence for the presence of other transits. 
We also conducted a grid search in phase to look for the secondary
eclipse with the transit period fixed.  We could not find a significant
detection of the secondary eclipse.  We put an 1-$\sigma$
upper limit of $65$\,ppm (0.065\,mmag) on the depth of the secondary 
eclipse.  This is estimated by computing the weighted average of the 
variance of points in-transit and the variance of 
points out-of-transit. The 1-$\sigma$ upper limit on the depth 
of the secondary eclipse we obtained from the {\em K2} \lc\ is much 
smaller than the predicted secondary eclipse depth (0.7\,mmag) 
if assuming the system is a blended binary 
(see Section \ref{sec:blend}).

\section{Analysis}
\label{sec:analysis}

\subsection{Stellar Parameters}
\label{sec:stellar}

\begin{figure}[]
\plotone{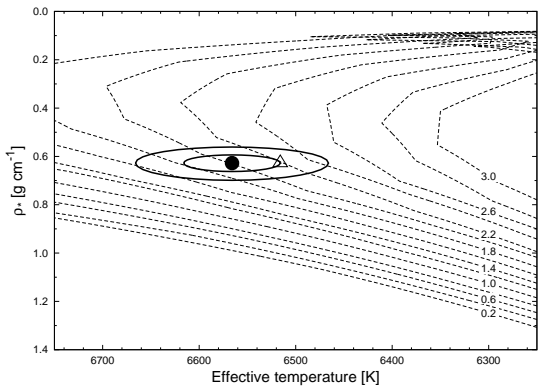}
\caption[]{
    Comparison between the measured values of \teffstar\ and
    \rhostar\ (filled circle), and the Y$^{2}$ model isochrones from
    \citet{yi:2001}. The best-fit values, and approximate 1$\sigma$
    and 2$\sigma$ confidence ellipsoids are shown. The open triangle
    shows the values from our initial SPC iteration. The Y$^{2}$
    isochrones are shown for ages of 0.2 to 3.0\,Gyr, in
    0.2\,Gyr increments.
\label{fig:iso}}
\end{figure}

The adopted stellar atmospheric parameters of \hatcur{} are
based on our SPC analysis of the 18 high S/N TRES spectra
(Section~\ref{sec:hispec} and Table~\ref{tab:spc}). SPC was applied
individually to each spectrum, and we take the error-weighted mean of
the individual atmospheric parameter measurements. The adopted
uncertainties ($0.1$\,dex for $\loggstar$, $50$\,K for \teffstar,
$0.080$\,dex for \feh\ and $0.50$\,\kms\ for \vsini) reflect our
estimate of the systematic errors in this method based on observations
of spectroscopic standard stars. Following \citet{sozzetti:2007} we
combine the resulting \teffstar\ and \feh\ measurements with the
\rhostar\ measurements from our joint analysis of the RV and light
curve data (Section~\ref{sec:globmod}), and compare them to the 
Yonsei-Yale stellar evolution models (\citealp{yi:2001}; the comparison is shown
in Figure~\ref{fig:iso}), to determine the physical stellar parameters
(mass, radius, age, luminosity, etc.).  As is often the case, we find
that the resulting $\loggstar$ value from this modeling differs
significantly from that determined through our initial SPC analysis,
and we therefore carried out a second iteration of SPC, fixing
\loggstar\ to the isochrone-based value. Repeating the joint RV+light
curve analysis and the stellar evolution look-up, we find that the
$\loggstar$ value had converged, and therefore did not carry out any
further iterations. The final parameters that we adopt for \hatcur{}
are listed in Table~\ref{tab:stellar} together with identifying
information and catalog photometry. We find that the star \hatcur{}
has a mass of \hatcurISOmlong{}\,\msun, a radius of
\hatcurISOrlong{}\,\rsun, an age of \hatcurISOage{}\,Gyr, and is at a
reddening-corrected distance of \hatcurXdistred{}\,pc (where we use
the \citealp{cardelli:1989} extinction law with $R_{V} = 3.1$).

\subsection{Global Modeling of RVs and Light Curves}
\label{sec:globmod}

We carried out a joint analysis of the TRES RVs and the HATNet,
KeplerCam and {\em K2} light curves following \cite{bakos:2010:hat11}
with modifications described by \cite{hartman:2012:hat39hat41}. The
RVs are modeled using a Keplerian orbit, while the light curves are
fit with a \citet{mandel:2002} transit model, with quadratic limb
darkening coefficients adopted from \citet{claret:2004} and
\citet{claret:2011}. The HATNet light curve that we analyzed has been
filtered via the EPD and TFA procedures before fitting the model, so
we include a dilution factor to account for possible over-filtering of
the transits. For the {\em K2} light curve, we integrate the model over
the 30\,min exposure time (this is done by evaluating the model flux
ratio at four evenly spaced times within a 30\,min bin, and taking the
average). We used equation~8 of \citet{kipping:2011:indkep} for our
{\em K2} model to account for a possible occultation signature and/or
a variation due to reflected light (in practice we find that the light
curve is consistent with no such variations). 
The {\em K2 } \lc\ has been fitted simultaneously with the TFA algorithm 
using 47 template stars observed in the same channel. 
For the KeplerCam light
curve, we use a simple model for instrumental trends (consisting of a
quadratic function of time and linear functions in the point spread
function (PSF) shape parameters), which we fit simultaneously with the
physical model. We use a differential evolution Markov-chain Monte
Carlo \citep[DEMCMC;][]{terbraak:2006} procedure to explore the
fitness landscape and to determine the posterior parameter
distributions.

The fit was performed both fixing the eccentricity to zero, and
allowing it to vary. We find that the TRES RVs do prefer a slight
eccentricity of $e = \hatcurRVecceneccen{}$, but based on the Bayesian
evidence (estimated from the Markov-chain following the method of
\citealp{weinberg:2013}), we conclude that the difference in $\chi^2$
is not significant enough to justify the additional free parameters,
and therefore we adopt the circular orbit model. The $95$\% confidence
upper limit on the eccentricity is $e
\hatcurRVeccentwosiglimeccen{}$. 

Table~\ref{tab:planetparam} lists the adopted parameters for the
planet \hatcurb{}. We find that this planet has a mass of
$\hatcurPPmlong{}$\,\mjup, a radius of $\hatcurPPrlong{}$\,\rjup, and
is orbiting its host star with a period of $\hatcurLCP$\,days and an
orbital separation of $\hatcurPParel$\,AU. At this separation the
planet would have an equilibrium temperature of $\hatcurPPteff$\,K
assuming zero albedo and complete redistribution of heat.

\subsection{Blend Analysis}
\label{sec:blend}

In order to rule out the possibility that \hatcur{} is a blended
stellar eclipsing binary system we carried out a blend analysis
following \citet{hartman:2012:hat39hat41}, with a few modifications 
to properly handle the {\em K2} light curve. These 
include: (1) integrating each simulated light curve model 
over the 30\,min exposures and using the
integrated model in calculating the $\chi^2$ difference from the
observations \citep{kipping:2011:indkep}; 
(2) using the \citet{claret:2011} limb darkening
coefficients for the {\em Kepler} band-pass; (3) using a polynomial
transformation from $griz$ to the {\em Kepler} $Kp$ magnitude system
to predict the relative fluxes of blended stars in the {\em Kepler}
band-pass.  

\begin{figure*}[]
\plotone{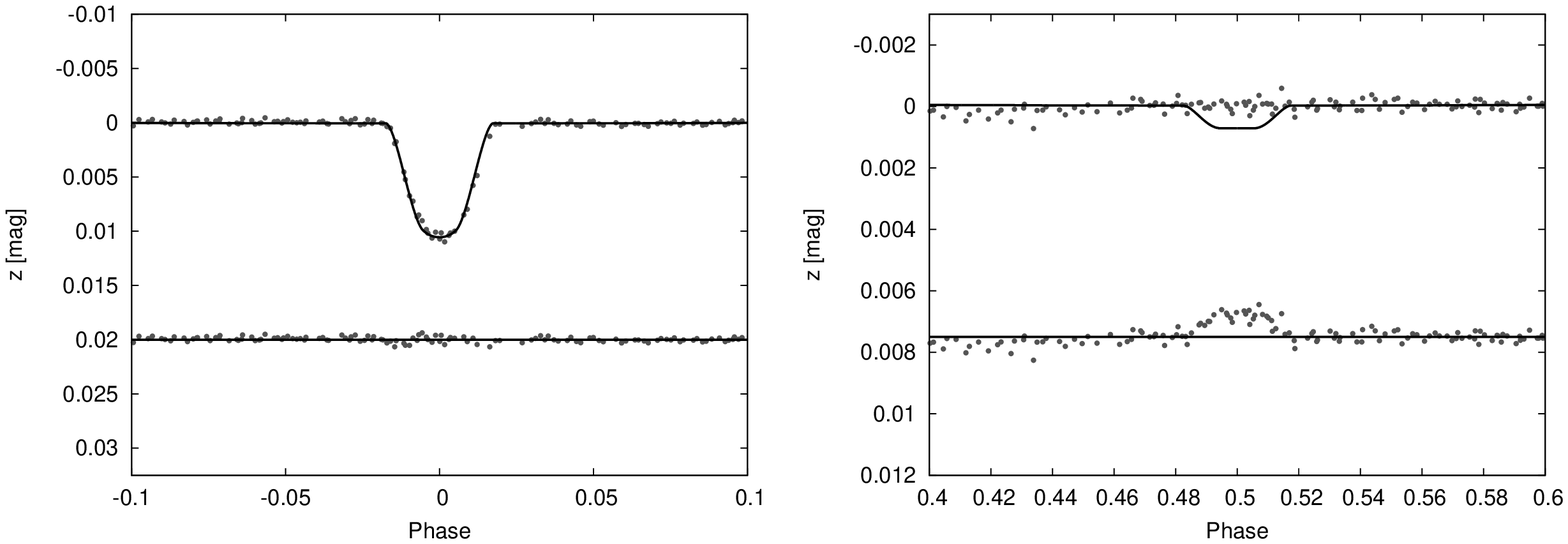}
\caption[]{
    Comparison between the phased {\em K2} observations of \hatcur{}
    and the predicted light curve for the best-fit blended stellar
    eclipsing binary model (top curve and data in each panel) together
    with the residuals from the model (bottom of each panel). In the
    left panel we show the primary eclipse, while in the right we show
    the secondary eclipse. The simulated primary eclipse fits the
    observed transit, but a significant secondary eclipse is
    predicted, but not observed. Note the different vertical scales in
    the two panels.
\label{fig:k2blend}}
\end{figure*}

We find that we can rule out a blended eclipsing binary scenario with
greater than $5\sigma$ confidence based solely on the photometry 
(including both HATNet and {\em K2} photometry). The
{\em K2} data is key in making this assessment. Although blend models
exist which fit the primary transit, such models predict a secondary
eclipse with a depth of $\Delta K_{p} = 0.7$\,mmag, which is ruled out
by the {\em K2} light curve (Figure~\ref{fig:k2blend}). Further
evidence against a blend scenario is the lack of BS variations (the
TRES BS measurements have an r.m.s.~scatter of 77\,\ms) and the
significant RV variation in phase with the photometric ephemeris and
consistent with a transiting planet.

We also considered the possibility that \hatcur{} is a transiting
planet system with a fainter, unresolved, stellar companion. From the
2MASS catalog there are no known stars within 20\arcsec\ of \hatcur{}
with $\Delta K < 5$\,mag. The highest spatial resolution observations
available are our KeplerCam observations which have a PSF FWHM of
3.5\arcsec. Based on these images
we can rule out a companion with $\Delta i \la 3.5$\,mag to within
5.4\arcsec. Based on our blend modeling we find that models including
a stellar companion with $M > 0.67$\,\msun\ yield a higher $\chi^2$
value than a single star with a transiting planet, but, except for
companions very close in mass to \hatcurb{}, which can be ruled out
with $\ga 3\sigma$ confidence, the difference in $\chi^2$ between
models with and without companions is not statistically
significant. Higher spatial resolution imaging, and/or long term RV
monitoring is needed to check for a binary star companion. If such a
companion exists, the mass and radius of \hatcurb{} would both be
larger than what we infer here.

\ifthenelse{\boolean{emulateapj}}{
  \begin{deluxetable*}{lcr}
}{
  \begin{deluxetable}{lcr}
}
\tablewidth{0pc}
\tabletypesize{\scriptsize}
\tablecaption{
    Stellar Parameters for \hatcur{} 
    \label{tab:stellar}
}
\tablehead{
    \multicolumn{1}{c}{~~~~~~~~Parameter~~~~~~~~} &
    \multicolumn{1}{c}{Value}                     &
    \multicolumn{1}{c}{Source}    
}
\startdata
\noalign{\vskip -3pt}
\sidehead{Identifying Information}
~~~~R.A. (h:m:s)                      &  \hatcurCCra{} & 2MASS\\
~~~~Dec. (d:m:s)                      &  \hatcurCCdec{} & 2MASS\\
~~~~GSC ID                            &  \hatcurCCgsc{} & GSC\\
~~~~2MASS ID                          &  \hatcurCCtwomass{} & 2MASS\\
~~~~EPIC ID                          &  \hatcurCCepic{} & EPIC\\
\sidehead{Spectroscopic properties}
~~~~$\teffstar$ (K)\dotfill         &  \hatcurSMEteff{} & SPC \tablenotemark{a}\\
~~~~$\meh$\dotfill                  &  \hatcurSMEzfeh{} & SPC                 \\
~~~~$\vsini$ (\kms)\dotfill         &  \hatcurSMEvsin{} & SPC                 \\
~~~~$\gamma_{\rm RV}$ (\kms)\dotfill&  35.11$\pm$0.1 & TRES                  \\
\sidehead{Photometric properties}
~~~~$B$ (mag)\dotfill               &  \hatcurCCtassmB{} & APASS                \\
~~~~$V$ (mag)\dotfill               &  \hatcurCCtassmv{} & TASS Mark IV               \\
~~~~$I$ (mag)\dotfill               &  \hatcurCCtassmI{} & TASS Mark IV   \\    ~~~~$Kep$ (mag)\dotfill               &  \hatcurCCkepmag{} & EPIC         \\
~~~~$g$ (mag)\dotfill               &  \hatcurCCtassmg{} & APASS                \\
~~~~$r$ (mag)\dotfill               &  \hatcurCCtassmr{} & APASS                \\
~~~~$i$ (mag)\dotfill               &  \hatcurCCtassmi{} & APASS                \\
~~~~$J$ (mag)\dotfill               &  \hatcurCCtwomassJmag{} & 2MASS           \\
~~~~$H$ (mag)\dotfill               &  \hatcurCCtwomassHmag{} & 2MASS           \\
~~~~$K_s$ (mag)\dotfill             &  \hatcurCCtwomassKmag{} & 2MASS           \\
\sidehead{Derived properties}
~~~~$\mstar$ ($\msun$)\dotfill      &  \hatcurISOmlong{} & Isochrones+\hatcurlumind{}+SPC \tablenotemark{b}\\
~~~~$\rstar$ ($\rsun$)\dotfill      &  \hatcurISOrlong{} & Isochrones+\hatcurlumind{}+SPC         \\
~~~~$\loggstar$ (cgs)\dotfill       &  \hatcurISOlogg{} & Isochrones+\hatcurlumind{}+SPC         \\
~~~~$\lstar$ ($\lsun$)\dotfill      &  \hatcurISOlum{} & Isochrones+\hatcurlumind{}+SPC         \\
~~~~$M_V$ (mag)\dotfill             &  \hatcurISOmv{} & Isochrones+\hatcurlumind{}+SPC         \\
~~~~$M_K$ (mag,\hatcurjhkfilset{})&  \hatcurISOMK{} & Isochrones+\hatcurlumind{}+SPC         \\
~~~~Age (Gyr)\dotfill               &  \hatcurISOage{} & Isochrones+\hatcurlumind{}+SPC         \\
~~~~$A_{V}$ (mag) \tablenotemark{c}\dotfill           &  \hatcurXAv{} & Isochrones+\hatcurlumind{}+SPC\\
~~~~Distance (pc)\dotfill           &  \hatcurXdistred{} & Isochrones+\hatcurlumind{}+SPC\\
~~~~\rhostar (\gcmc)\dotfill & \hatcurISOrholong{} & LC \tablenotemark{d} \\
\enddata
\tablenotetext{a}{
    SPC = ``Stellar Parameter Classification'' method based on
    cross-correlating high-resolution spectra against synthetic
    templates \citep{buchhave:2012:spc}. These parameters rely primarily
    on SPC, but have a small dependence also on the iterative analysis
    incorporating the isochrone search and global modeling of the
    data, as described in the text.  } 
\tablenotetext{b}{
    Isochrones+\hatcurlumind{}+SPC = Based on the Y$^{2}$ isochrones
    \citep{yi:2001},
    the stellar density used as a luminosity indicator, and the SPC
    results.
} 
\tablenotetext{c}{ 
Total \band{V} extinction to the star determined
  by comparing the catalog broad-band photometry listed in the table
  to the expected magnitudes from the
  Isochrones+\hatcurlumind{}+SPC model for the star. We use the
  \citet{cardelli:1989} extinction law.  
}
\tablenotetext{d}{
The stellar density is determined primarily from fitting the transit 
light curve. There is also a slight dependency on the SPC atmospheric 
parameters and stellar evolution models through the adopted limb 
darkening coefficients.
}
\ifthenelse{\boolean{emulateapj}}{
  \end{deluxetable*}
}{
  \end{deluxetable}
}
\ifthenelse{\boolean{emulateapj}}{
  \begin{deluxetable*}{lc}
}{
  \begin{deluxetable}{lc}
}
\tabletypesize{\scriptsize}
\tablecaption{Parameters for the transiting planet \hatcurb{}.\label{tab:planetparam}}
\tablehead{
    \multicolumn{1}{c}{~~~~~~~~Parameter~~~~~~~~} &
    \multicolumn{1}{c}{Value \tablenotemark{a}}                     
}
\startdata
\noalign{\vskip -3pt}
\sidehead{\Lc{} parameters}
~~~$P$ (days)             \dotfill    & $\hatcurLCP{}$              \\
~~~$T_c$ (${\rm BJD}$)    
      \tablenotemark{b}   \dotfill    & $\hatcurLCT{}$              \\
~~~$T_{14}$ (days)
      \tablenotemark{b}   \dotfill    & $\hatcurLCdur{}$            \\
~~~$T_{12} = T_{34}$ (days)
      \tablenotemark{b}   \dotfill    & $\hatcurLCingdur{}$         \\
~~~$\arstar$              \dotfill    & $\hatcurPPar{}$             \\
~~~$\zrstar$ \tablenotemark{c}              \dotfill    & $\hatcurLCzeta{}$\phn       \\
~~~$\rpl/\rstar$          \dotfill    & $\hatcurLCrprstar{}$        \\
~~~$b^2$                  \dotfill    & $\hatcurLCbsq{}$            \\
~~~$b \equiv a \cos i/\rstar$
                          \dotfill    & $\hatcurLCimp{}$           \\
~~~$i$ (deg)              \dotfill    & $\hatcurPPi{}$\phn         \\

\sidehead{Limb-darkening coefficients \tablenotemark{d}}
~~~$c_1,i$ (linear term)  \dotfill    & $\hatcurLBii{}$            \\
~~~$c_2,i$ (quadratic term) \dotfill  & $\hatcurLBiii{}$           \\
~~~$c_1,r$               \dotfill    & $\hatcurLBir{}$             \\
~~~$c_2,r$               \dotfill    & $\hatcurLBiir{}$            \\
~~~$c_1,Kep$               \dotfill    & $\hatcurLBikep{}$             \\
~~~$c_2,Kep$               \dotfill    & $\hatcurLBiikep{}$            \\

\sidehead{RV parameters}
~~~$K$ (\ms)              \dotfill    & $\hatcurRVK{}$\phn\phn      \\
~~~$e$ \tablenotemark{e}  \dotfill    & $\hatcurRVeccentwosiglimeccen{}$ \\
~~~RV jitter (\ms) \tablenotemark{f}        \dotfill    & $<$ 15           \\

\sidehead{Planetary parameters}
~~~$\mpl$ ($\mjup$)       \dotfill    & $\hatcurPPmlong{}$          \\
~~~$\rpl$ ($\rjup$)       \dotfill    & $\hatcurPPrlong{}$          \\
~~~$C(\mpl,\rpl)$
    \tablenotemark{g}     \dotfill    & $\hatcurPPmrcorr{}$         \\
~~~$\rhopl$ (\gcmc)       \dotfill    & $\hatcurPPrho{}$            \\
~~~$\log g_p$ (cgs)       \dotfill    & $\hatcurPPlogg{}$           \\
~~~$a$ (AU)               \dotfill    & $\hatcurPParel{}$          \\
~~~$T_{\rm eq}$ (K) \tablenotemark{h}        \dotfill   & $\hatcurPPteff{}$           \\
~~~$\Theta$ \tablenotemark{i} \dotfill & $\hatcurPPtheta{}$         \\
~~~$\langle F_j \rangle$ ($10^{9}$\ergscmsq) \tablenotemark{i}
                          \dotfill    & $\hatcurPPfluxavg{}$       \\ [-1.5ex]
\enddata
\tablenotetext{a}{
    The adopted parameters assume a circular orbit. Based on the
    Bayesian evidence ratio we find that this model is strongly
    preferred over a model in which the eccentricity is allowed to
    vary in the fit. For each parameter we give the median value and
    68.3\% (1$\sigma$) confidence intervals from the posterior
    distribution.
}
\tablenotetext{b}{
    Reported times are in Barycentric Julian Date calculated directly
    from UTC, {\em without} correction for leap seconds.
    \ensuremath{T_c}: Reference epoch of mid transit that
    minimizes the correlation with the orbital period.
    \ensuremath{T_{14}}: total transit duration, time
    between first to last contact;
    \ensuremath{T_{12}=T_{34}}: ingress/egress time, time between first
    and second, or third and fourth contact.
}
\tablenotetext{c}{
    Reciprocal of the half duration of the transit used as a jump
    parameter in our MCMC analysis in place of $\arstar$. It is
    related to $\arstar$ by the expression $\zrstar = \arstar
    (2\pi(1+e\sin \omega))/(P \sqrt{1 - b^{2}}\sqrt{1-e^{2}})$
    \citep{bakos:2010:hat11}.
}
\tablenotetext{d}{
    Values for a quadratic law, adopted from the tabulations by
    \cite{claret:2004,claret:2013} according to the spectroscopic (SPC) parameters
    listed in \reftabl{stellar}.
}
\tablenotetext{e}{
    The 95\% confidence upper-limit on the eccentricity from a model
    in which the eccentricity is allowed to vary in the fit.
}
\tablenotetext{f}{
    Error term, either astrophysical or instrumental in origin, added
    in quadrature to the formal RV errors. This term is varied in the
    fit assuming a prior inversely proportional to the jitter. 
    In this case we find a preferred value of $0$ for the jitter, 
    and list the 95\% confidence upper limit.
}
\tablenotetext{g}{
    Correlation coefficient between the planetary mass \mpl\ and
    radius \rpl\ determined from the parameter posterior distribution
    via $C(\mpl,\rpl) = <(\mpl - <\mpl>)(\rpl -
    <\rpl>)>/(\sigma_{\mpl}\sigma_{\rpl})>$ where $< \cdot >$ is the
    expectation value operator, and $\sigma_x$ is the standard
    deviation of parameter $x$.
}
\tablenotetext{h}{
    Planet equilibrium temperature averaged over the orbit, calculated
    assuming a Bond albedo of zero, and that flux is reradiated from
    the full planet surface.
}
\tablenotetext{i}{
    The Safronov number is given by $\Theta = \frac{1}{2}(V_{\rm
    esc}/V_{\rm orb})^2 = (a/\rpl)(\mpl / \mstar )$
    \citep[see][]{hansen:2007}.
}
\tablenotetext{j}{
    Incoming flux per unit surface area, averaged over the orbit.
}
\ifthenelse{\boolean{emulateapj}}{
  \end{deluxetable*}
}{
  \end{deluxetable}
}

\subsection{Out of transit variation of \hatcur{}}

\begin{figure}[]
\plotone{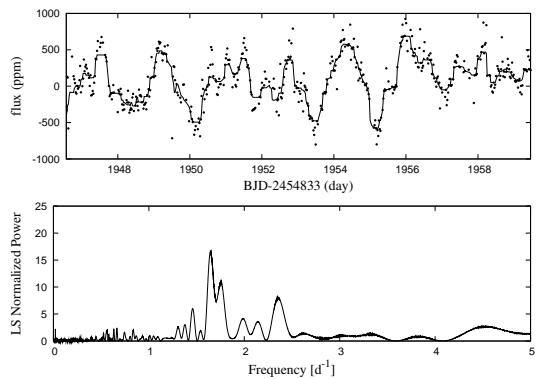}
\caption[]{
    The top panel shows a segment of the {\em K2} \lc{} of \hatcur, 
    demonstrating the out of transit variability pattern. 
    The bottom panel shows the Lomb-Scargle normalized power 
    spectrum of the {\em K2} \lc{} excluding the transits.
\label{fig:oot}}
\end{figure}

Stellar variability with amplitude $\sim$1\, mmag are present 
in the {\em K2} \lc{} (see Figure \ref{fig:oot}). The Lomb-Scargle 
normalized power spectrum \citep{lomb:1976,scargle:1982} 
of the star suggest that it is likely a pulsating $\gamma$ Dor, 
with the primary period of $\sim 1.644 \pm0.03$ days. The second 
strongest period peak is at $\sim 1.744\pm0.023$ days.  
The effective temperature and the surface gravity of the star 
place it outside the low temperature boundary of the classical 
instability strip, but within the range of other 
$\gamma$ Dors discovered by the {\em Kepler Mission} 
\citep{uytterhoeven:2011}.

We tried to constrain the orbital phase variation due to \hatcurb{} 
with the K2 \lc. To take into account the influence of the stellar 
variability, we simultaneously fit for the amplitude of the six 
most dominant Fourier modes detected by the Lomb-Scargle algorithm, 
together with the expected reflection, beaming, and ellipsoid 
variation effects with periods constrained due to the presence 
of the planet. The best fitted values and error bars are presented in 
Table \ref{tab:phase}. We report a 3-$\sigma$ detection of the 
reflection effect, with 
an amplitude of $21\pm7$\,ppm, which is comparable to the theoretically 
estimated value ($27.4$\,ppm, following \citet{mazeh:2010}). 
We do not detect the beaming effect. 
The ellipsoidal variation is detected with a 2-$\sigma$ significance. 
However, we caution the reader, that 
we also obtain a high amplitude coefficient for the 
$\sin{(2\pi/(P_{orb}/2)t_j)}$ term, which indicates the detected 
ellipsoidal variation is not in phase with the planet transit. 
This could be due to our poor understanding of the stellar 
variability, suggesting the modeled ellipsoidal variation amplitude 
may not be physically meaningful.

\section{Discussion}
\label{sec:discussion}

In this paper we presented the discovery and characterization 
of \hatcurb, an inflated massive hot Jupiter around a bright 
F star.

The radius anomaly of hot Jupiters is one of the oldest unsolved 
problems in the exoplanet field \citep{baraffe:2010,spiegel:2012}. 
Observationally, only a few very inflated planets ($R>1.35R_J$)  
are observed to be above the mass of $2M_J$.
It has been empirically demonstrated by various authors that more 
massive planets are harder to inflate. 
\citet{enoch:2012} derived the dependence of giant planet radius 
in three mass regimes, and found that the radii of 
high mass planets ($M>2M_J$) are less sensitive to the equilibrium 
temperature of the planets, as compared to Jupiter mass planets.  
\citet{weiss:2013} fitted for the fundamental plane of giant planets 
using existing data. They found that the radius of irradiated hot 
Jupiters inversely correlates with the planet mass. 
\citet{zhou:2014} investigated the mass dependence of planet radius 
on the equilibrium temperature, and also found that as the planet mass 
increase above $1M_J$, the influence on radius from irradiation 
decreases. 

\begin{figure}[!ht]
\epsscale{1.1}
\plotone{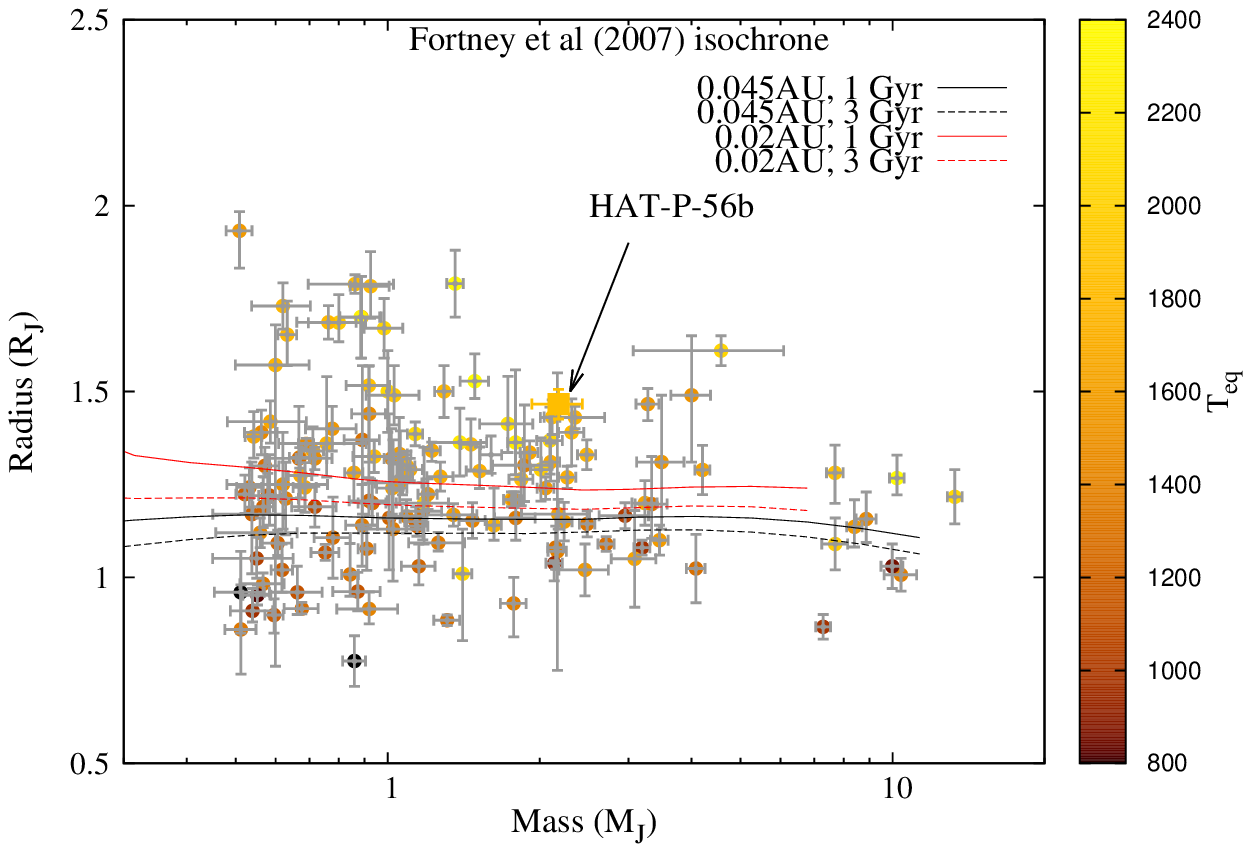}
\caption{
  Mass-radius diagram for \hatcurb{} (big square) compared to 
  the full sample of confirmed hot Jupiters (circles) from 
  http://exoplanets.org. The points 
  are color coded with their equilibrium temperature (cooler planets 
  have darker colors). The error bars indicate one sigma 
  uncertainties. The lines represents the 
  \citet{fortney:2007} planet radius model without 
  heavy elements cores. The black lines are for planets at 0.045 AU from 
  a sun like star (solid:1\,Gyr, dashed:3\,Gyr). 
  The red lines are for planets at 0.02 AU from 
  a sun like star (solid:1\,Gyr, dashed:3\,Gyr).
\label{fig:MR}
}
\end{figure}

\begin{figure}[!ht]
\epsscale{1.1}
\plotone{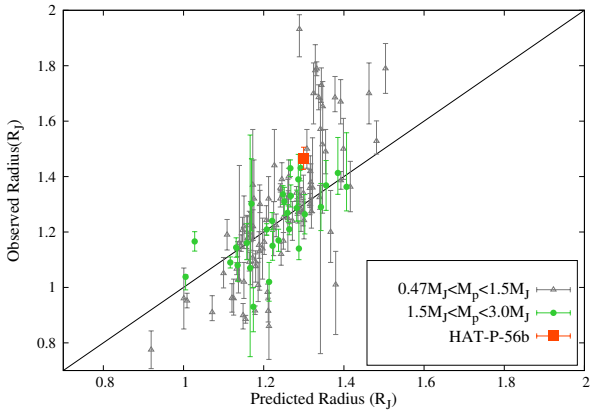}
\caption{
  Observed radius of hot Jupiters versus the radius predicted by the 
  empirical relation of \citep{weiss:2013}. The green filled 
  circles represent planets with mass $1.5M_J<M_p<3.0M_J$; the 
  light grey open triangles represent planets with 
  mass $0.47<M_p<1.5M_J$. The red square is \hatcurb. 
  \label{fig:Rplane}}
\end{figure}

In Figure \ref{fig:MR} we show the location of \hatcurb{} in the mass 
radius diagram of hot Jupiters. Taken at face value, it is the most 
inflated hot Jupiter with a mass between 1.5\mjup$-$4\mjup. 
We show the position of \hatcurb{} relative to the fundamental 
plane fitted by \citet{weiss:2013} in Figure \ref{fig:Rplane}, 
where it can be seen 
that \hatcurb{} is the farthest away from the fundamental plane when 
compared to the other massive planets (green points in the figure). 
The fundamental plane of irradiated hot Jupiters 
(M$>$150\mearth, or 0.47\mjup), is expressed as the following 
\citep{weiss:2013}: 
\begin{equation*}
\frac{R_p}{\rearth} = 2.45 (\frac{M_P}{\mearth})^{-0.039}(\frac{F}{ergs^{-1}cm^{-2}})^{0.094}.
\end{equation*}
The unusually large radius of \hatcurb, given its mass, makes it an 
important data point in the hot Jupiter population.

However, it is noteworthy that the radius of \hatcurb{} may have a 
large uncertainty due to the large impact parameter 
($b = \hatcurLCimp{}$) indicating a nearly grazing transit. 
The grazing transit geometry makes the transit depth highly 
dependent on the limb darkening parameters of the star. The 
radius uncertainties are largely reduced thanks to the high 
precision {\em K2} \lc. 
Another way to better constrain the radius of the planet would be to
observe transits in the near infrared where limb darkening is
negligible.
The only other planets known to be on close-to grazing orbits 
are TrES-2b, HAT-P-27b, WASP-34b, WASP-67b,
and Kepler-447b 
\citep{odonovan:2006,beky:2011:hat27,hellier:2012,
mancini:2014,lillo-box:2015:kepler447b}. While the grazing
configuration increases the uncertainty of the planetary radius
measurement, as \cite{ribas:2008} 
pointed out, such a near grazing transit has the advantage of 
its depth and duration being more sensitive to the presence of other
planetary companions on inclined orbits. This makes \hatcurb{} a promising 
target among the hot Jupiters for detecting transit timing and duration 
variations.

Given the brightness of the host star ($V_{\rm mag}=10.9$), 
\hatcur{} is a system of particular interest for measuring the 
spin-orbit obliquity angle. The estimated RM effect amplitude 
$\Delta\,V_R$ is $\sim$ 140\ms\ following Equation 5 of 
\citet{gaudiwinn:2007}. \hatcur{} is a relatively fast 
rotating F star with effective temperature (\hatcurSMEteff\,K) 
slightly above the \citet{albrecht:2012} division for tidal 
alignment of hot Jupiters. Measurement of the spin-orbit obliquity 
will contribute an important data 
point to the obliquity distribution statistics.
The rotation of \hatcur{} is similar to CoRoT-11, and rapid enough
that the distortion in the spectral line profiles during transit may
be resolved, enabling transit Doppler tomography of the system 
\citep{gandolfi:2012,colliercameron:2010}. Similar to CoRoT-11, 
the rotation period of \hatcur{} is faster than the orbit 
period of \hatcurb{}, which is unusual \citep{walkowicz:2013}. This 
may either suggest that the tidal interaction between the star and 
planet is weak, or the planet might be pushed out by the tides of 
the star from a closer orbit.

\ifthenelse{\boolean{emulateapj}}{
  \begin{deluxetable*}{lccc}
}{
  \begin{deluxetable}{lccc}
}
\tablewidth{0pt}
\tabletypesize{\scriptsize}
\tablecaption{
    Derived and Expected Orbit Phase Variation \citep{mazeh:2010} for \hatcur{} \lc\ 
    \label{tab:phase}
}
\tablehead{
    \multicolumn{1}{c}{~~~~~~~~Coefficient~~~~~~~~} &
    \multicolumn{1}{c}{Derived Value } &
    \multicolumn{1}{c}{Expected Value \tablenotemark{a}} &    
    \multicolumn{1}{c}{Effect} \\
    \multicolumn{1}{c}{} &
    \multicolumn{1}{c}{(ppm)} &
    \multicolumn{1}{c}{(ppm)\tablenotemark{a}} &    
    \multicolumn{1}{c}{} 
}
\startdata
\noalign{\vskip -3pt}
~~~~a$_1c$                       & $-21_{-7}^{+7}$& $-27.4\pm0.6$ & Reflection\\
~~~~a$_1s$                       &$1.6_{-7}^{+7}$ & $3.5\pm0.4$ & Beaming \\
~~~~a$_2c$                       & $-17_{-7}^{+7}$ &$-6.5\pm0.8$ & Ellipsoidal\\
~~~~a$_2s$                       & $45_{-7}^{+7}$ &- & - \\
\enddata
\tablenotetext{a}{
The expected values are computed following \citet{mazeh:2010}.
} 
\ifthenelse{\boolean{emulateapj}}{
  \end{deluxetable*}
}{
  \end{deluxetable}
}



\acknowledgements 

\paragraph{Acknowledgements}
We thank the referee for his/her useful comments.
HATNet operations have been funded by NASA grants NNG04GN74G and
NNX13AJ15G. Follow-up of HATNet targets has been partially supported
through NSF grant AST-1108686. G.\'A.B., Z.C. and K.P. acknowledge
partial support from NASA grant NNX09AB29G.  K.P. acknowledges support
from NASA grant NNX13AQ62G. We acknowledge partial support also from the 
{\em Kepler Mission} under NASA Cooperative Agreement NNX13AB58A (D.W.L.,
PI). Data presented in this paper are based on observations 
obtained at the HAT station at the Submillimeter Array of SAO, 
and the HAT station at the Fred Lawrence Whipple Observatory 
of SAO. The authors wish to recognize and acknowledge the 
very significant cultural role and reverence that the summit of 
Mauna Kea has always had within the indigenous Hawaiian community. 
We are most fortunate to have the opportunity to conduct 
observations from this mountain. This paper presents observations 
made with the Apache Point Observatory 
3.5-meter telescope, which is owned and operated by the 
Astrophysical Research Consortium. The {\em K2} data presented in this 
paper were obtained from the Mikulski Archive for Space 
Telescopes (MAST). STScI is operated by the Association of 
Universities for Research in Astronomy, Inc., under NASA 
contract NAS5-26555. Support for MAST for non-HST data is 
provided by the NASA Office of Space Science via grant NNX09AF08G 
and by other grants and contracts. This paper includes data 
collected by the Kepler telescope. 
Funding for the {\em K2 Mission} is provided by the NASA Science Mission directorate.
\clearpage

\end{document}